# Glass-like Cross-plane Thermal Conductivity of Kagome Metals RbV$_3$Sb$_5$ and CsV$_3$Sb$_5$

Yu Pang[1], Jinjin Liu[2,5], Zeyu Xiang[1], Xuanhui Fan[3], Jie Zhu[3,*], Zhiwei Wang[2,5,6,*], Yugui Yao[2,5,6], Xin Qian[1,*],Ronggui Yang[1,4,*]

[1] School of Energy and Power Engineering, Huazhong University of Science and Technology, Wuhan 430074, China

[2] Centre for Quantum Physics, Key Laboratory of Advanced Optoelectronic Quantum Architecture and Measurement (MOE), School of Physics, Beijing Institute of Technology, Beijing 100081, China

[3] School of Energy and Power Engineering, Dalian University of Technology, Dalian 116024, China

[4] State Key Laboratory of Coal Combustion, Huazhong University of Science and Technology, Wuhan 430074, China

[5] Beijing Key Lab of Nanophotonics and Ultrafine Optoelectronic Systems, Beijing Institute of Technology, Beijing 100081, China

[6] Material Science Center, Yangtze Delta Region Academy of Beijing Institute of Technology, Jiaxing 314011, China

*Corresponding emails: zhujie@dlut.edu.cn; zhiweiwang@bit.edu.cn; xinqian21@hust.edu.cn; ronggui@hust.edu.cn

## Abstract

This work reports the thermal conductivity of RbV$_3$Sb$_5$ and CsV$_3$Sb$_5$ with three-dimensional charge density wave phase transitions from 80 K to 400 K measured by pump-probe thermoreflectance techniques. At room temperature, the in-plane (basal plane) thermal conductivities are found moderate, i.e., 12 W m$^{-1}$ K$^{-1}$ of RbV$_3$Sb$_5$ and 8.8 W m$^{-1}$ K$^{-1}$ of CsV$_3$Sb$_5$, and ultralow cross-plane (stacking direction) thermal conductivities are observed, with 0.72 W m$^{-1}$ K$^{-1}$ of RbV$_3$Sb$_5$ and 0.49 W m$^{-1}$ K$^{-1}$ of CsV$_3$Sb$_5$ at 300 K. A unique glass-like temperature dependence in the cross-plane thermal conductivity is discovered, which decreases monotonically even lower than the Cahill-Pohl limit as the temperature decreases below the phase transition point $T_{\text{CDW}}$. This temperature dependence is found to obey the hopping transport picture. In addition, a peak in cross-plane thermal conductivity is observed at $T_{\text{CDW}}$ as a fingerprint of the modulated structural distortion along the stacking direction.

Two-dimensional kagome lattice is a topological system with flat electronic bands, Dirac cones, and Van Hove singularities[1,2], which induce many intriguing physical phenomena such as spin liquid states[3], bond density wave order[4], superconductivity[5] and charge density wave (CDW)[6]. The recently discovered layered kagome metals $AV_3Sb_5$ (A = K, Rb, Cs)[7] with $\mathbb{Z}_2$-type nontrivial band topology[8,9] provide a unique platform for studying electron correlations, topological effects, and quantum phase transition[10,11]. Particularly, $AV_3Sb_5$ exhibits lots of interesting phenomena, such as anomalous Hall effects[12], spontaneous symmetry breaking[13], competition between CDW with superconductivity[14–16], etc.

The CDW in $AV_3Sb_5$ is featured by simultaneous lattice distortions both in each basal plane and the modulated stacking of different distortion patterns[17,18]. Recent measurements on magnetization, heat capacity, and electrical resistivity identify that $AV_3Sb_5$ undergoes a first-order CDW phase transition at the critical temperatures $T_{CDW} \approx$ 78-102 K[8,9,19]. Above the CDW transition ($T>T_{CDW}$), the crystal structure of $AV_3Sb_5$ consists of a high-symmetry vanadium-based kagome sublattice, as shown in Fig. 1a-b. However, density functional theory (DFT) study predicts that the high-symmetry structure exhibits phonon instability below $T_{CDW}$, as softening acoustic phonon modes emerge near $M$ and $L$ points in the Brillouin zone[20]. The soft phonon modes drive the vanadium atoms in each kagome layer spontaneously shift away from the high-symmetry sites, forming 2×2 superlattices with the star of David (SD) or inverse star of David (ISD) patterns[20]. More interestingly, DFT study further predicts that lattice distortion happens not only inside the basal planes but such distortion is modulated with a stacking period of 2 or 4 layers[18,20]. The 2×2 in-plane superlattice is observed earlier using both XRD[8] and scanning tunneling microscopy (STM)[21–23]. Recently, a variety of experimental measurements further confirm multiple superstructures of 3D CDW ordering, such as 2×2×2[17], 2×2×4[18], or even both[24], etc. Despite these discoveries, origin of the 3D CDW in $AV_3Sb_5$ remains elusive. A dilemma is that inelastic X-ray scattering shows the absence of clear phonon softening or Kohn anomaly[25], which is contradictory to DFT predictions[20]. In addition, Subires *et al.* recently proposed that CDW phase change in $AV_3Sb_5$ is of order-disorder type which goes beyond the conventional weak-coupling regime of Peierls transitions[26].

Characterizing thermal properties could provide a unique angle to study the mechanisms of CDW phase transition. For example, thermal conductivity shows a peak or a sudden drop near $T_{CDW}$. In CDW materials with one-dimensional (1D) lattice distortions such as $K_{0.3}MoO_3$ and $(TaSe_4)_2I$[27], a peak in temperature-dependent thermal conductivity is usually observed near the phase transition point $T_{CDW}$, due to the excess heat carried by phasons or amplitudons as quantized modes of CDW[28]. In layered materials such as 1T-TaS$_2$, and 2H-TaSe$_2$[29,30] with two-dimensional (2D) lattice distortions, strong electron-phonon interactions driving the phase transition are also manifested in a sudden drop in thermal conductivity near $T_{CDW}$ due to the additional scattering of heat-carrying phonons. Till now, 3D CDW has only been identified in a few systems such as $YBa_2Cu_3O_{6.67}$[31] and 1T-VSe$_2$[32] other than $AV_3Sb_5$, while how 3D lattice distortions affect thermal transport remains largely unexplored. Recently, Yang *et al.* reported a glass-like in-plane thermal conductivity in $AV_3Sb_5$ due to charge fluctuations above $T_{CDW}$[33], but measurements of cross-plane thermal conductivity near the CDW transition are still in urgent need for a complete understanding of thermal transport in layered $AV_3Sb_5$ with 3D CDW.

In this work, temperature-dependent thermal conductivity along the cross-plane direction of kagome metals ($RbV_3Sb_5$ and $CsV_3Sb_5$) are measured using the pump-probe thermoreflectance technique[34]. This work observes that $AV_3Sb_5$ exhibits a unique glass-like temperature dependence in the cross-plane thermal conductivity, where the thermal conductivity drops rapidly with decreasing temperatures lower than the Cahill-Pohl limit[35] at temperatures below $T_{CDW}$. A hopping model is proposed to explain this trend. In addition, a peak at the $T_{CDW}$ is observed in the temperature-dependent cross-plane thermal conductivity. Our observation of the glass-like cross-plane thermal conductivity aligns with the recent discovery that the CDW transition in $AV_3Sb_5$ is three-dimensional and is of order-disorder type beyond the conventional weak-coupling regime[17,26].

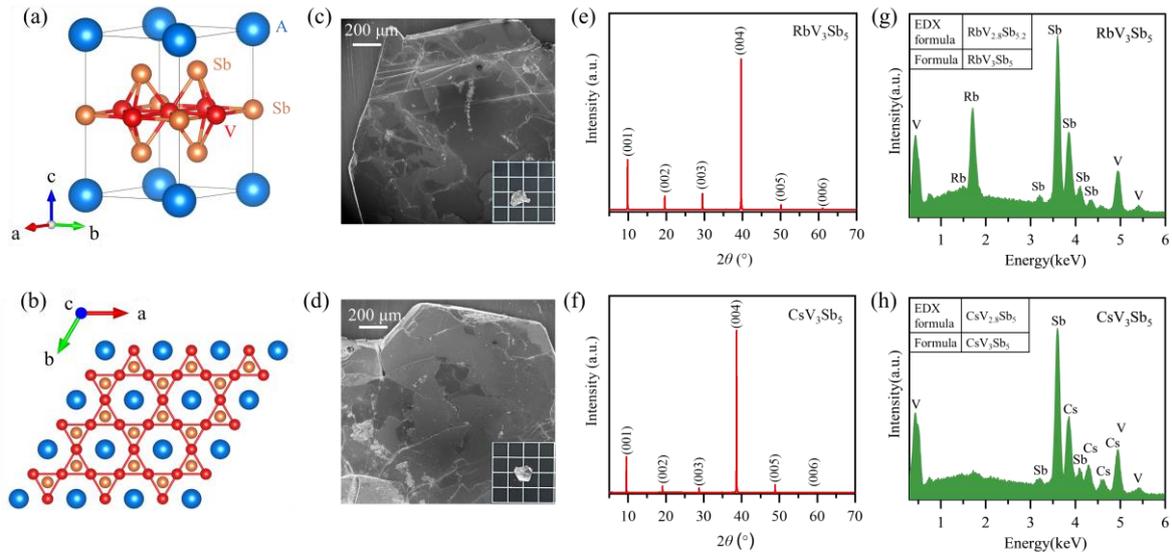

Fig. 1. (a) Unit cell of AV$_3$Sb$_5$, (A = K, Rb, Cs) (b) atomic structure of AV$_3$Sb$_5$ projected to the (001) plane. A, V, and Sb atoms are presented as blue, red, and orange balls, respectively. SEM and optical (inset) images of (c) RbV$_3$Sb$_5$ and (d) CsV$_3$Sb$_5$. Insets show the optical image of the corresponding materials, and each grid in the optical image is 3 mm by 3 mm. XRD and EDX characterization of (e-g) RbV$_3$Sb$_5$ and (f-h) CsV$_3$Sb$_5$.

Single crystals of RbV$_3$Sb$_5$ and CsV$_3$Sb$_5$ are prepared with binary Rb-Sb and Cs-Sb fluxes using the self-flux method[36]. Alkali metals (Rb and Cs, Alfa Aesar, 99.8%), vanadium pieces (Aladdin, 99.97%), and Sb (Alfa Aesar, 99.9999%) with a molar ratio of 9: 3: 17 are loaded into an alumina crucible and sealed in an evacuated quartz tube. These raw materials are heated at 5 K/min and kept at 1273 K for 24 h. After that, the ampoules are cooled down to 473 K at 3 K/h. The residual flux is removed using deionized water. The morphology of the synthesized samples is characterized by scanning electron microscope (SEM), as shown in Fig. 1c-d. XRD characterizations are shown in Fig. 1e-f. The full width at half maxima (FWHM) of the diffraction peaks on the (00*l*) crystal plane of the two samples are both nearly 0.07°, indicating good crystallinity of the synthesized samples. Energy dispersive X-ray spectroscopy (EDX) results suggest that the samples crystallize in RbV$_{2.8}$Sb$_{5.2}$ (Fig. 1g) and CsV$_{2.8}$Sb$_5$ (Fig. 1h), respectively, which agree well with nominal compositions (inset, Figs. 1g and 1h).

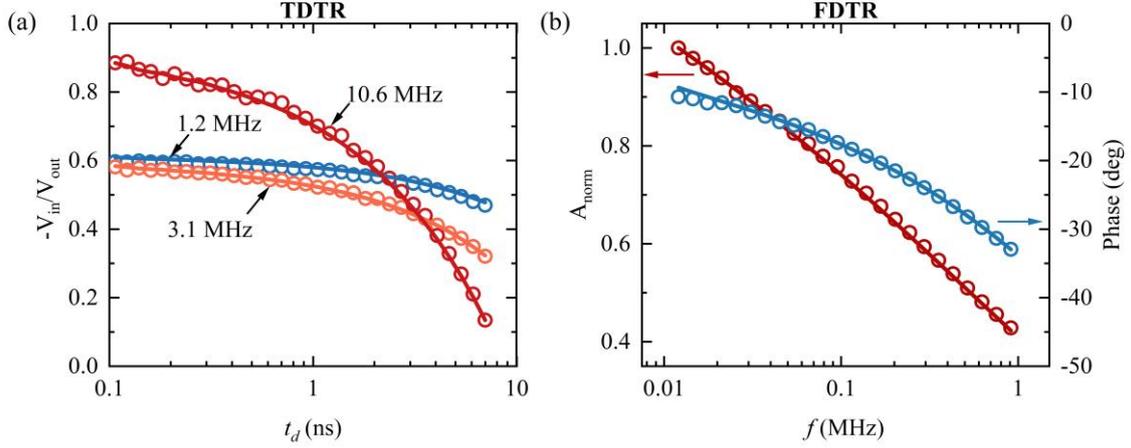

Fig. 2. (a) TDTR and (b) FDTR experimental signals (open symbols) at room temperature along with the fitting curves calculated based on the heat transfer model. The TDTR signals -$V_{in}/V_{out}$ are acquired using an identical large beam size ($w$ = 20 μm) and different modulation frequencies $f$ = 1.2, 3.1, and 10.6 MHz. The FDTR data were taken from low modulation frequency range $f$ = 0.1-1 MHz, with the $A_{norm}$ and phase signals being both well-fitted.

Time-domain thermoreflectance (TDTR)[37] and frequency-domain thermoreflectance (FDTR) techniques[38] are used to measure the room-temperature cross-plane and in-plane thermal conductivity ($\kappa_z$ and $\kappa_r$) of AV$_3$Sb$_5$, respectively. Fig. 2a shows typical TDTR signals of the CsV$_3$Sb$_5$ sample using a root-mean-square spot radius of 20 μm and multiple modulation frequencies of 1.2 to 10.6 MHz. In this case, the heat transfer is quasi-one-dimensional in the cross-plane direction, and the TDTR signal is only sensitive to $\kappa_z$. The cross-plane thermal conductivity $\kappa_z$ and the interface conductance $G$ are the fitting parameters in data reduction. We found that a single set of $\kappa_z$ and $G$ can achieve nice fitting of signals at different frequencies, suggesting no modulation frequency dependence originated from non-Fourier phonon transport or local nonequilibrium effects[39]. The $\kappa_z$ is determined as 0.49 ± 0.07 W m$^{-1}$ K$^{-1}$ at room temperature using TDTR with large spot radius (~ 20 μm). After $\kappa_z$ is deteremined, in-plane thermal conductivity $\kappa_r$ is measured as 8.8 ± 1.5 W m$^{-1}$ K$^{-1}$ using FDTR technique. A tightly focused spot radius of 3 μm and low modulation frequency range $f$ = 0.01-1 MHz is used when performing FDTR measurements to ensure high sensitivity to $\kappa_r$, as shown in Fig. 2b. Similarly, $\kappa_z$ and $\kappa_r$ of RbV$_3$Sb$_5$ at room temperature are also measured as 0.72 ± 0.10 W m$^{-1}$ K$^{-1}$ and 12 ± 2 W m$^{-1}$ K$^{-1}$, respectively. Our measurement of $\kappa_r$ is consistent with the recently report value[12,33]. More details about pump-probe measurement, sensitivity analysis, and uncertainty analysis are listed in Supplemental Materials S1-3.

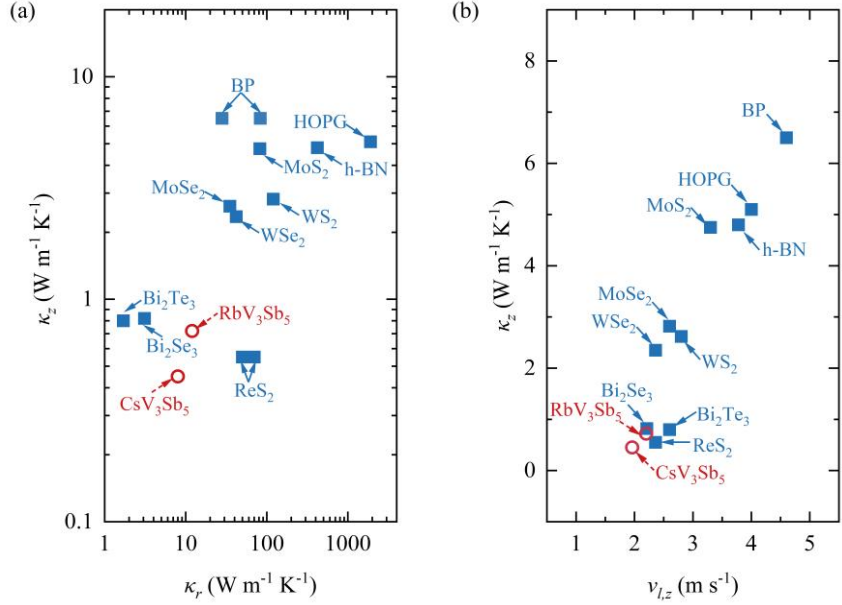

Fig. 3. (a) Summary of experimentally measured cross-plane thermal conductivity $\kappa_z$ and in-plane thermal conductivity $\kappa_r$ of layered crystalline materials at room temperature. (b) Collected cross-plane thermal conductivity $\kappa_z$ of layered crystalline materials versus longitudinal sound velocity $v_{l,z}$. The cross-plane thermal conductivities come from BP[40], HOPG[41], h-BN[42], transition metal dichalcogenides (MoS$_2$, WS$_2$, MoSe$_2$, WSe$_2$)[39], ReS$_2$[43], and bismuth dichalcogenides (Bi$_2$Te$_3$, Bi$_2$Se$_3$)[44,45], as well as kagome metals AV$_3$Sb$_5$ (A = Rb, Cs) measured in this work. The longitudinal sound velocity $v_{l,z}$ is obtained through phonon dispersion or experimentally determined data. The data of longitudinal sound velocities are taken from BP[40], HOPG[46], h-BN[42], transition metal dichalcogenides (MoS$_2$, WS$_2$, MoSe$_2$, WSe$_2$)[47], ReS$_2$[48], bismuth dichalcogenides (Bi$_2$Te$_3$, Bi$_2$Se$_3$)[49,50], and kagome metals AV$_3$Sb$_5$ (A = Rb, Cs)[20].

We compare room-temperature $\kappa_z$ and $\kappa_r$ of AV$_3$Sb$_5$ in Fig. 3a, as well as other typical van der Waals materials. Among the materials gathered, AV$_3$Sb$_5$ exhibits the lowest $\kappa_r$ other than bismuth dichalcogenides (Bi$_2$Te$_3$, Bi$_2$Se$_3$), while CsV$_3$Sb$_5$ even displays the lowest $\kappa_z$. Fig. 3b summarizes the longitudinal sound velocity in the cross-plane and measured $\kappa_z$ of collected materials. The $v_{l,z}$ of AV$_3$Sb$_5$, which are 1960 m s$^{-1}$ and 2200 m s$^{-1}$ for CsV$_3$Sb$_5$ and RbV$_3$Sb$_5$, respectively, are also relatively low against other layered materials (e.g., $v_{l,z}$ = 2360 m s$^{-1}$ for ReS$_2$[43] and $v_{l,z}$ = 2210 m s$^{-1}$ for Bi$_2$Se$_3$[49]). The low $v_{l,z}$ indicates that the low $\kappa_z$ of AV$_3$Sb$_5$ at room temperature is mainly caused by the weak interlayer bonding strength.

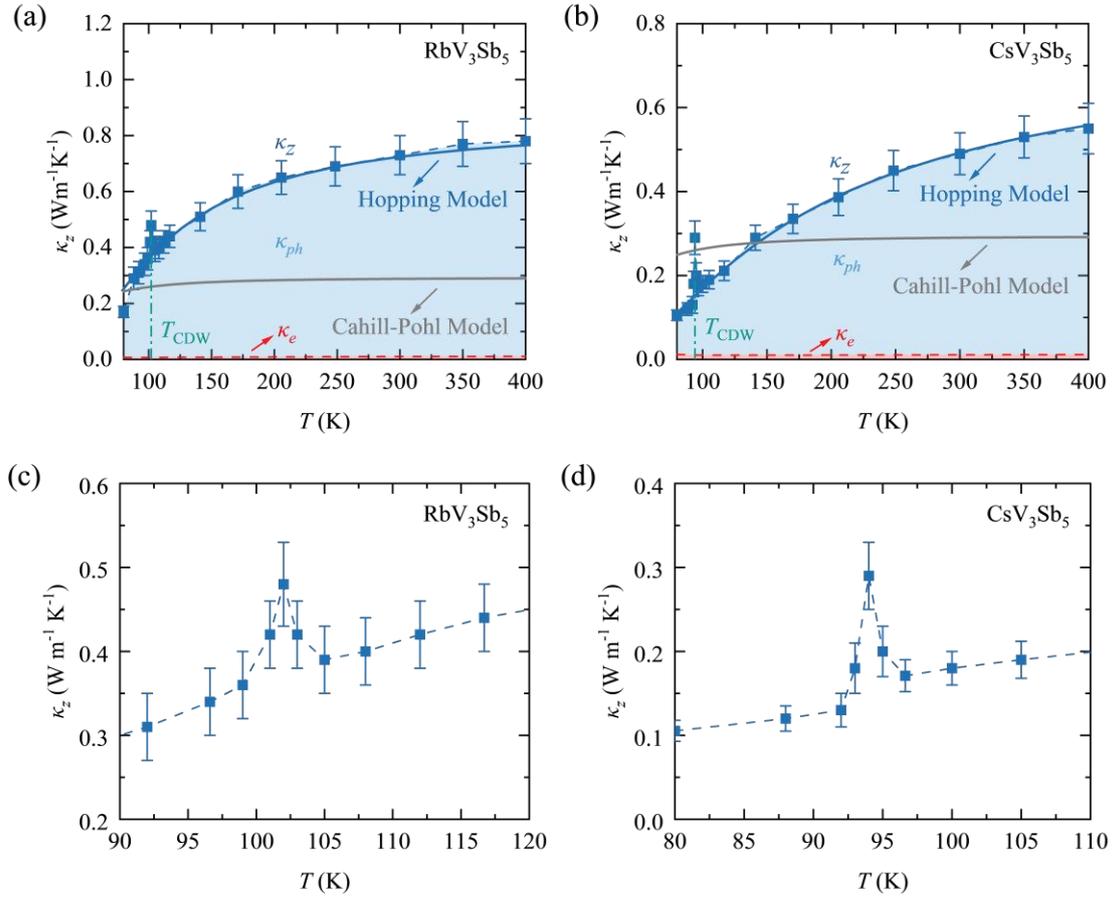

Fig. 4. Measured cross-plane thermal conductivity (a) $\kappa_z$ of RbV$_3$Sb$_5$ and (b) $\kappa_z$ of CsV$_3$Sb$_5$, over the temperature range of 80-400 K, as well as electron/phonon-contributed terms. Thermal conductivity due to electrons $\kappa_e$ is obtained using the Wiedemann-Franz law, with phonon thermal conductivity $\kappa_{ph}$ acquired through subtracting $\kappa_e$ from the measured thermal conductivity $\kappa$. Discrete points, red dotted lines, and middle coloring sections represent the $\kappa$, $\kappa_e$, and $\kappa_{ph}$, respectively, together with cyan dot-dash line lines used for highlighting the $T_{CDW}$. The blue solid lines are estimated phonon thermal conductivity along cross-plane direction based on the hopping model. The gray solid lines are lower limit to thermal conductivity estimated by the Cahill-Pohl model. (c-d) Details for the peak in $\kappa_z$ around $T_{CDW}$ for (c) RbV$_3$Sb$_5$ and (d) CsV$_3$Sb$_5$.

Figure 4a-b show the temperature-dependent $\kappa_z$ (80-400 K) of RbV$_3$Sb$_5$ and CsV$_3$Sb$_5$, as well as the electron and phonon contributions estimated using the Wiedemann-Franz law (see Supplementary Materials S4). Our measurement shows that electron contribution is negligible in the cross-plane direction, which is typical for layered materials due to the weak interlayer coupling and low charge mobilities in the cross-plane direction[51]. Interestingly, the cross-plane

thermal conductivity $\kappa_z$ shows a glass-like temperature dependence with thermal conductivity increasing monotonically with the temperature. The thermal conductivity even drops below the Cahill-Pohl limit below $T_{\text{CDW}}$[35]. Our observation is aligned with the observation that localized phonon modes drive structural fluctuations near $T_{\text{CDW}}$[26]. Considering these localized modes could be well-described by Einstein oscillators, we proposed a hopping model to describe the cross-plane thermal transport. In general, the lattice thermal conductivity of crystalline solids can be expressed as a frequency-dependent integral over the phonon spectrum:

$$\kappa_{ph}(T) = \int C(\omega)D(\omega)d\omega \tag{1}$$

where $C(\omega) = \hbar\omega g(\omega)\frac{\partial f}{\partial T}$ is the spectral volumetric specific heat, with $f$ the Bose-Einstein distribution and $g(\omega)$ the density of states, and $D(\omega)$ is spectral diffusivity for each vibration. In this work, the dispersion of propagating low-frequency phonons is described by the Callaway model[52], while localized high-frequency modes are treated as Einstein oscillators. The $D(\omega)$ of propagating modes is described as $v(\omega)^2\tau(\omega)/3$ using the phonon-gas model, where $v$ is group velocity and $\tau$ is relaxation time. For the localized modes, the transport process is described by a hopping model, with hopping length $\alpha$, hopping rate expressed as $\omega/\pi$ and an Arrhenius-like probability of successful hopping $P(T) \propto \exp(-E_a/k_B T)$, with $E_a$ denoting the activation energy scale. The $D(\omega)$ of the localized modes is thus expressed as:

$$D(\omega) = \frac{\alpha^2 \omega}{\pi} P(T) \tag{2}$$

Integrating over the vibrational spectra, the lattice thermal conductivity based on the hopping model is written as:

$$\kappa_{ph}(T) = k_B(n - n_E)\sum_j v_j^2 \left(\frac{T}{\theta_{c_j}}\right)^3 \int_0^{\frac{\theta_{c_j}}{T}} \frac{x^4 e^x}{(e^x - 1)^2}\tau(x,T)dx \\ + 3n_E k_B \frac{x_E^2 e^x}{(e^{x_E} - 1)^2}\alpha^2 \frac{\pi}{\omega_E} P(T) \tag{3}$$

where $x = \hbar\omega/k_B T$, $\omega_c$ is the cut-off frequency between the propagating phonons and localized modes, $n$ is number density of atoms, $n_E$ is the number density of Einstein oscillators, $v_j$ is the group velocity of acoustic branch $j$. $\theta_{cj}$ is similar to the Debye temperature of branch $j$, defined as $\theta_{cj} = \hbar\omega_{cj}/k_B$. $\omega_E$ denotes the average frequency of Einstein oscillators. More details of the hopping model are listed in Supplemental Material S5. Fig. 4c-d show that the hopping model well captures the glass-like temperature dependence of $\kappa_z$. Our measurement also showed a peak (Figs. 4c and 4d) at $T_{CDW}$ in temperature-dependent $\kappa_z$, which supports recent identifications of 3D CDW lattice distortions[17,18]. Based on the proposed hopping model, the excess $\kappa_z$ contributed by the CDW phase change can be estimated from the peak heights, with 0.08 W m$^{-1}$ K$^{-1}$ and 0.14 W m$^{-1}$ K$^{-1}$ for RbV$_3$Sb$_5$ and CsV$_3$Sb$_5$, respectively.

In summary, this work systematically measures the temperature-dependent thermal conductivity of the kagome metals AV$_3$Sb$_5$ (A = Rb, Cs) using pump-probe thermoreflectance techniques. The two kagome metals exhibit moderate in-plane thermal conductivities at room temperature, with 12 W m$^{-1}$ K$^{-1}$ of RbV$_3$Sb$_5$ and 8.8 W m$^{-1}$ K$^{-1}$ of CsV$_3$Sb$_5$. The cross-plane thermal conductivities are determined as 0.72 W m$^{-1}$ K$^{-1}$ and 0.49 W m$^{-1}$ K$^{-1}$ for RbV$_3$Sb$_5$ and CsV$_3$Sb$_5$, respectively. The low $\kappa_z$ arises from strongly reduced phonon group velocity associated with weak interlayer bonding. Unique glass-like temperature dependence of $\kappa_z$ is observed where a phonon hopping transport model is proposed. Our measurement is consistent with recent discoveries that the CDW phase transition in AV$_3$Sb$_5$ is of order-disorder type. Sudden increase in $\kappa_z$ has also been experimentally observed at $T_{CDW}$, supporting recent identifications of 3D CDW transitions where the lattice distortions are modulated in the cross-plane direction. The measurements of the thermal conductivities of AV$_3$Sb$_5$ serve as an important benchmark for understanding thermal transport in material systems with order-disorder phase transitions and dynamic instabilities.

**Acknowledgments**

X.Q. and R.Y. acknowledge support from National Key R & D Project from Ministry of Science and Technology of China (Grant No. 2022YFA1203100) and National Natural Science Foundation of China (NSFC Grant No. 52276065). Z.W acknowledges support from the National Natural Science Foundation of China (Grant No. 92065109), the National Key R&D Program of China (Grant Nos. 2020YFA0308800, 2022YFA1403401), the Beijing Natural Science Foundation (Grant Nos. Z190006, Z210006). J.Z. thanks support from National Natural Science Foundation of China (NSFC Grant No. 51976025). Y.P. acknowledges the support from Cross-Discipline Ph.D. Education Plan. Z. W. thanks the Analysis & Testing Center at BIT for assistance in facility support. The authors declare no conflict of interest.

# References

1. Ye, L. et al. Massive Dirac fermions in a ferromagnetic kagome metal. *Nature* **555**, 638–642 (2018).

2. Kang, M. et al. Dirac fermions and flat bands in the ideal kagome metal FeSn. *Nat. Mater.* **19**, 163–169 (2020).

3. Yan, S., Huse, D. A. & White, S. R. Spin-Liquid Ground State of the *S* = 1/2 Kagome Heisenberg Antiferromagnet. *Science* **332**, 1173–1176 (2011).

4. Isakov, S. V., Wessel, S., Melko, R. G., Sengupta, K. & Kim, Y. B. Hard-Core Bosons on the Kagome Lattice: Valence-Bond Solids and Their Quantum Melting. *Phys. Rev. Lett.* **97**, 147202 (2006).

5. Ko, W.-H., Lee, P. A. & Wen, X.-G. Doped kagome system as exotic superconductor. *Phys. Rev. B* **79**, 214502 (2009).

6. Guo, H. & Franz, M. Topological insulator on the kagome lattice. *Phys. Rev. B* **80**, 113102 (2009).

7. Ortiz, B. R. et al. New kagome prototype materials: discovery of $KV_3Sb_5$, $RbV_3Sb_5$, and $CsV_3Sb_5$. *Phys. Rev. Mater.* **3**, 094407 (2019).

8. Ortiz, B. R. et al. $CsV_3Sb_5$: A Z-2 Topological Kagome Metal with a Superconducting Ground State. *Phys. Rev. Lett.* **125**, 247002 (2020).

9. Ortiz, B. R. et al. Superconductivity in the Z-2 kagome metal $KV_3Sb_5$. *Phys. Rev. Mater.* **5**, 034801 (2021).

10. Zhao, H. et al. Cascade of correlated electron states in the kagome superconductor $CsV_3Sb_5$. *Nature* **599**, 216–221 (2021).

11. Fu, Y. et al. Quantum Transport Evidence of Topological Band Structures of Kagome Superconductor $CsV_3Sb_5$. *Phys. Rev. Lett.* **127**, 207002 (2021).

12. Zhou, X. et al. Anomalous thermal Hall effect and anomalous Nernst effect of $CsV_3Sb_5$. *Phys. Rev. B*

# Supplementary Materials: Glass-like Cross-plane Thermal Conductivity of Kagome Metals RbV$_3$Sb$_5$ and CsV$_3$Sb$_5$

Yu Pang[1], Jinjin Liu[2,5], Zeyu Xiang[1], Xuanhui Fan[3], Jie Zhu[3,*], Zhiwei Wang[2,5,6,*], Yugui Yao[2,5,6], Xin Qian[1,*], Ronggui Yang[1,4,*]

[1] School of Energy and Power Engineering, Huazhong University of Science and Technology, Wuhan 430074, China

[2] Centre for Quantum Physics, Key Laboratory of Advanced Optoelectronic Quantum Architecture and Measurement (MOE), School of Physics, Beijing Institute of Technology, Beijing 100081, China

[3] School of Energy and Power Engineering, Dalian University of Technology, Dalian 116024, China

[4] State Key Laboratory of Coal Combustion, Huazhong University of Science and Technology, Wuhan 430074, China

[5] Beijing Key Lab of Nanophotonics and Ultrafine Optoelectronic Systems, Beijing Institute of Technology, Beijing 100081, China

[6] Material Science Center, Yangtze Delta Region Academy of Beijing Institute of Technology, Jiaxing 314011, China

*Corresponding emails: zhujie@dlut.edu.cn; zhiweiwang@bit.edu.cn; xinqian21@hust.edu.cn; ronggui@hust.edu.cn

## Contents

## S1. Anisotropic thermal conductivity measurements

We combine TDTR and FDTR to simultaneously obtain $\kappa_z$ and $\kappa_r$ of kagome metals $AV_3Sb_5$ (A = Rb, Cs)[1,2] (Figure S1), where TDTR is used to determine $\kappa_z$ and FDTR is used to determine $\kappa_r$. The root-mean-square spot radius $w$ are 20 μm (TDTR) and 3 μm (FDTR), obtained by fitting the in-phase signal as a function of offset distances between the pump and the probe beams[3]. We use the four-point probe to characterize the electric conductivity of metal transducers, and the thermal conductivity is then estimated using the Wiedemann-Franz law. The thickness $h$ of Au and Al transducers is measured by the DektakXT profilometer (Bruker, USA). The temperature-dependent volumetric heat capacity $C$ of $AV_3Sb_5$ is taken from the reported theoretical fittings based on Physical Property Measurement System (PPMS) measurements[4]. Table S1 summarizes the parameters of the Al/$AV_3Sb_5$ and Au/$AV_3Sb_5$ samples for TDTR and FDTR measurements at room temperature, respectively.

Table S1. Parameters for TDTR/FDTR measurements at room temperature.

| Layer | $\kappa_z$ (W m$^{-1}$ K$^{-1}$) | $\kappa_r$ (W m$^{-1}$ K$^{-1}$) | $h$ (nm) | $w$ (μm) | $C$ (MJ m$^{-3}$ K$^{-1}$) | $G$ (MW m$^{-2}$ K$^{-1}$) |
|---|---|---|---|---|---|---|
| Al | 150 | 150 | 109 | 20 | 2.44 | 60 |
| Au | 220 | 220 | 100 | 3 | 2.49 | 30 |
| RbV$_3$Sb$_5$ | 0.72 | 8.8 | - | - | 1.47 | - |
| CsV$_3$Sb$_5$ | 0.49 | 12 | - | - | 1.50 | - |

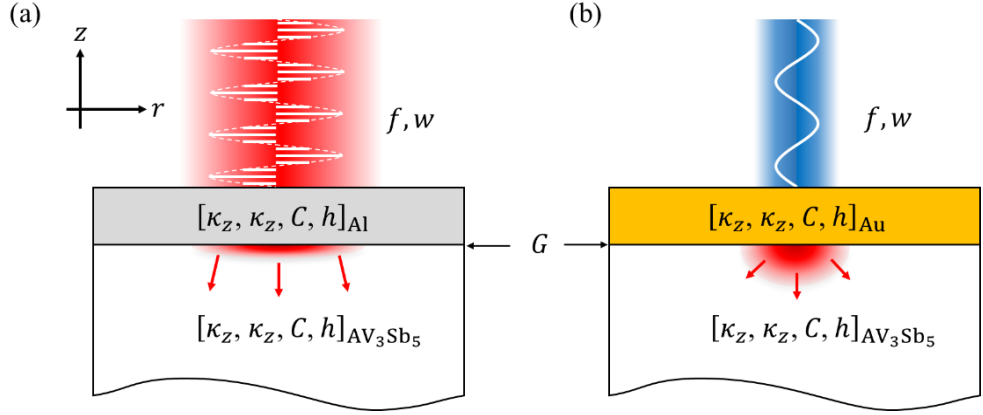

Fig. S1. Schematic of measuring the anisotropic thermal conductivity of $AV_3Sb_5$ using (a) TDTR and (b) FDTR.

## S2. Sensitivity analysis

We perform sensitivity analysis to determine experimental settings (modulation frequency range, laser spot radius, etc.) for characterizing anisotropic thermal conductivity. The sensitivity in pump-probe measurement is defined as:

$$S_x^y = \frac{\partial \ln y}{\partial \ln x} \quad (S1)$$

where $y$ denotes the signal (amplitude or phase) and $x$ represents the parameter of interest. Figure. S2 shows sensitivity analysis of $CsV_3Sb_5$ using TDTR and FDTR measurements with different experiment setups at 300 K. Parameters for computing experimental sensitivities are shown in Table S1. Fig. S2a-b shows the sensitivity curves of the TDTR system implemented in this work. With a root-mean-square spot radius $w$ of 20 μm, the TDTR measurement is primarily sensitive to $\kappa_z$ but insensitive to $\kappa_r$. Fig. S2c-d shows that the TDTR system implemented in our lab has low sensitivity to the in-plane thermal conductivity, even with a tightly focused laser spot with a radius of 4 μm and a modulation frequency of 1.2 MHz[5]. We, therefore, use the FDTR with the lowest modulation frequency of 10 kHz and a small $w$ of 3 μm to achieve high sensitivity to the in-plane thermal conductivity of $AV_3Sb_5$, as shown in Fig. S2e-f.

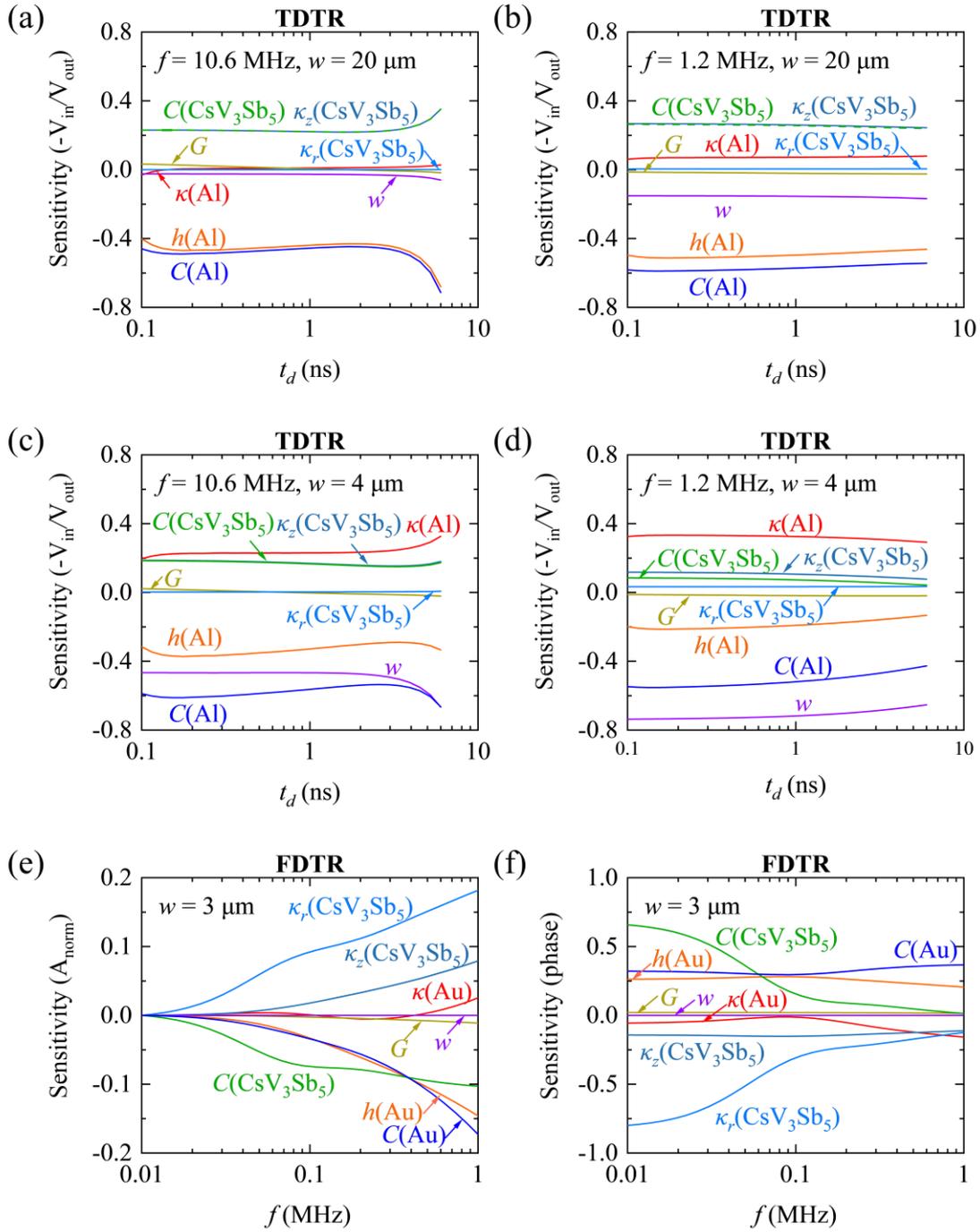

Fig. S2. Sensitivities of CsV$_3$Sb$_5$ using TDTR or FDTR measurements with different experiment setups: TDTR using -V$_{in}$/V$_{out}$ signal (a) with $f$ = 10.6 MHz and $w$ = 20 μm, (b) with $f$ = 1.2 MHz and $w$ = 20 μm, (c) with $f$ = 10.6 MHz and $w$ = 4 μm, (d) with $f$ = 1.2 MHz and $w$ = 4 μm; FDTR using (e) A$_{norm}$ and (f) phase signal with $f$ = 0.1-1 MHz and $w$ = 3 μm. Sensitivities of $\kappa_z$(CsV$_3$Sb$_5$) are higher than >0.2 in TDTR measurement using a large beam size, together with $\kappa_r$(CsV$_3$Sb$_5$) fully suppressed less than 0.01 (sensitivity at $f$ = 3.1 MHz is consistent but not shown here for clarity).

## S3. Uncertainty analysis

We use the multivariate error propagation formula based on Jacobi matrices for uncertainty analysis[6,7]. The variance of unknown parameters $X_U$ is calculated as:

$$\text{Var}[X_U] = (J_U^T J_U)^{-1} J_U^T (\text{Var}[y] + J_P \text{Var}[X_P] J_P^T) J_U (J_U^T J_U)^{-1} \qquad (S2)$$

where the superscript $T$ represents the matrix transpose, $J_U$ and $J_P$ are the Jacobian matrices correlating the signal $y$ and the unknown parameters $X_U$ or control parameters $X_P$, respectively. In this work, the control parameters $X_P$ include the root-mean-square radius $w$, the properties of the metal transducers ($\kappa$, $C$, and $h$), and the heat capacity of AV$_3$Sb$_5$. Typical uncertainty levels are: 10% for $\kappa(\text{Al})/\kappa(\text{Au})$, 3% for $C(\text{Al})/C(\text{Au})$ and $C(\text{AV}_3\text{Sb}_5)$, 4% for $h(\text{Al})/h(\text{Au})$, and 5% for $w$. For TDTR measurements, the vector of unknown parameters $X_U$ include $\kappa_z$ of AV$_3$Sb$_5$ and the interface conductance $G$. After TDTR measurements are performed, $\kappa_z$ is included as the control parameter $X_P$ when performing uncertainty analysis of FDTR, and $X_U$ of FDTR only includes $\kappa_r$ and $G$. Table S2 summarizes the $\text{Var}[X_U]$ for TDTR and FDTR fitting of CsV$_3$Sb$_5$ at 300 K. The uncertainties ($2\sigma$) of each unknown parameter can be calculated as the square root of the diagonal elements of the matrix $\text{Var}[X_U]$.

Table. S2. Covariance matrices $\text{Var}[X_U]$ of the experimental data measured upon TDTR and FDTR for CsV$_3$Sb$_5$ at 300 K, fitted using the multilayered heat transfer model. The units are $\kappa_z$ (W m$^{-1}$ K$^{-1}$), $\kappa_r$ (W m$^{-1}$ K$^{-1}$) and $G$ (MW m$^{-2}$ K$^{-1}$).

| TDTR | $G$ | $\kappa_z$ |
|---|---|---|
| $G$ | 176 | -0.129 |
| $\kappa_z$ | -0.129 | 0.0049 |
| Best-fit | 60 | 0.49 |
| Uncertainty | 21.8 % | 14.3 % |
| FDTR | $G$ | $\kappa_r$ |
| $G$ | 2600 | -27.4 |
| $\kappa_r$ | -27.4 | 2.3 |
| Best-fit | 30 | 8.8 |
| Uncertainty | 174% | 17.1% |

## S4. Estimation of $\kappa_e$ using Wiedemann-Franz law

The electron thermal conductivity $\kappa_e$ and phonon thermal conductivity $\kappa_{ph}$ are obtained using the Wiedemann-Franz law $\kappa_e = LT/\rho$ where $L$ and $\rho$ are Lorenz number and electrical resistivity, respectively. More specifically, in-plane electrical resistivity $\rho_r$ of AV$_3$Sb$_5$ is first acquired using a physical property measurement system (PPMS), over a temperature range of 80-300 K. The high-temperature (300-400 K) $\rho_r$ is estimated through linear extrapolation of the measured $\rho_r$ above Debye temperature (175.5 K for RbV$_3$Sb$_5$, 142 K for CsV$_3$Sb$_5$ [8]), in which $\rho_r$ is linearly dependent on temperature [9], as shown in as shown in Fig. S3a. Cross-plane electrical resistivity $\rho_z$ is subsequently estimated using the formula $\rho_z = \alpha \rho_r$, where $\alpha$ is the ratio between $\rho_z$ and $\rho_r$. The $\alpha$ has been reported to be about 600 and nearly temperature-independent[10]. Once $\kappa_e$ along the in-plane and cross-plane directions (shown in Fig. S3b-c) are determinate from $\rho$, phonon thermal conductivity $\kappa_{ph}$ is acquired by subtracting $\kappa_e$ from the measured thermal conductivity $\kappa$.

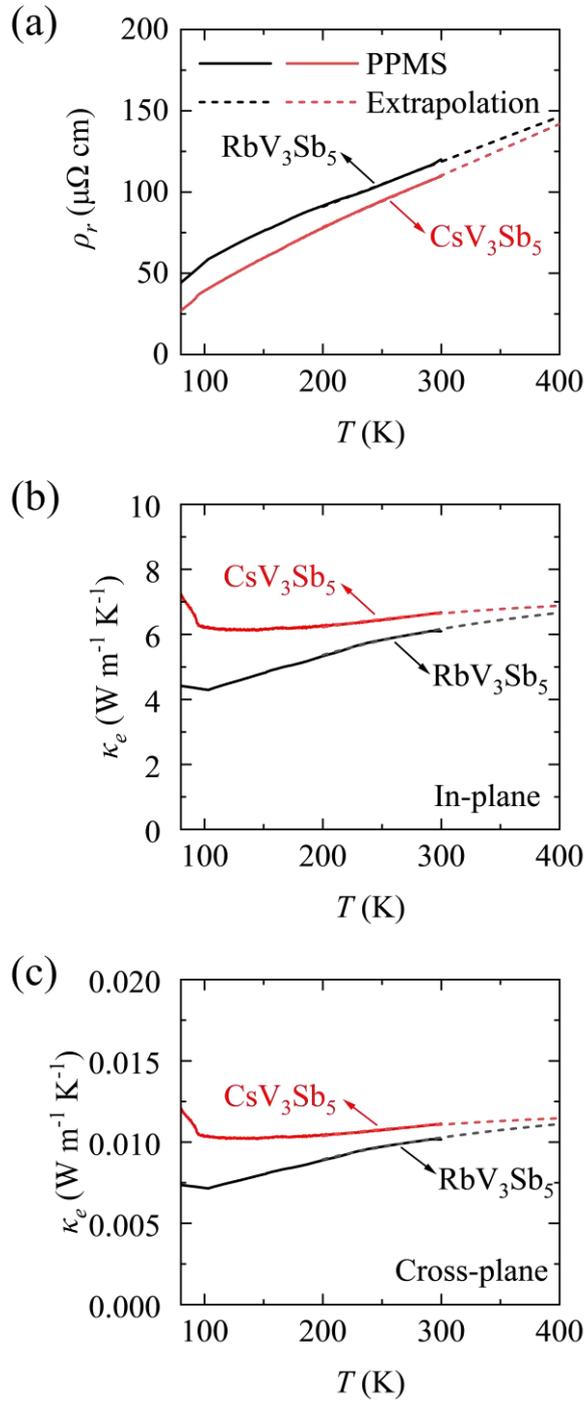

Fig. S3. (a) in-plane electrical resistivity, (b) in-plane electron thermal conductivity, and (c) cross-plane electron thermal conductivity of $AV_3Sb_5$ (A = Rb, Cs).

## S5. Hopping model for cross-plane thermal transport

In order to describe the glass-like cross-plane thermal transport, a hopping model is proposed in this work, inspired by the recent observation of localized vibration modes driving the order-disorder CDW phase transition in AV$_3$Sb$_5$[11]. Our hopping model separately treats the propagating low-frequency phonons and the high-frequency localized modes. The low-frequency phonons are described by the Callaway model with the Debye approximation, while the localized high-frequency modes can be treated as Einstein oscillators. Therefore, the vibrational density of states is written as:

$$g(\omega) = \begin{cases} \dfrac{3\omega^2}{2\pi^2 v_s^3}, & \omega < \omega_c \\ 3n_E \delta(\omega - \omega_E), & \omega > \omega_c \end{cases} \quad (S3)$$

where $\omega_c$ is the cut-off frequency between the propagating phonons and localized modes, $v_s$ is the averaged acoustic velocity, and $n_E$ denotes the number density of Einstein oscillators. With the Debye approximation, the total number of propagating modes below $\omega_c$ should equal $3(n - n_E)$, therefore:

$$\int_0^{\omega_c} g(\omega) d\omega = 3(n - n_E) \quad (S4)$$

which results in an explicit expression correlating $n_E$ and $\omega_c$:

$$\omega_c = [6\pi^2 v_s^3 (n - n_E)]^{1/2} \quad (S5)$$

where $n$ is the number density of atoms. Following Allen-Feldman's definition of spectral thermal diffusivity $D(\omega)$, the thermal conductivity can be computed by integrating $C(\omega)D(\omega)$ over the vibrational spectrum. For low-frequency propagating phonons, $D(\omega)$ is simply $v(\omega)^2 \tau(\omega)/3$ from the phonon-gas model, with $\tau(\omega)$ denoting the relaxation time. The propagation of the localized modes can be described by a random walk theory[12], where the $D(\omega)$ is determined by the hopping length $\alpha$, hopping rates expressed as $\omega/\pi$ [13], and probability of a successful hopping $P$. Considering the structural fluctuation is driven by finite temperature, we phenomenologically describe the probability of successful hopping $P$ using the

Arrhenius law $P = Ae^{-\frac{E_a}{k_B T}}$, where $E_a$ is activation energy and $A$ is the preexponential factor. Therefore, the $D(\omega)$ over the phonon spectrum is expressed as:

$$D(\omega) = \begin{cases} \dfrac{1}{3} v_s^2 \tau(\omega), & \omega < \omega_c \\ \alpha^2 \dfrac{\pi}{\omega} A e^{-\frac{E_a}{k_B T}}, & \omega > \omega_c \end{cases} \tag{S5}$$

Substituting Eq. S(3-5) into Eq. (3) of the main text, we obtain calculated lattice thermal conductivity:

$$\kappa_{ph}(T) = k_B (n - n_E) \sum_j v_j^2 \left(\frac{T}{\theta_{cj}}\right)^3 \int_0^{\frac{\theta_{cj}}{T}} \frac{x^4 e^x}{(e^x - 1)^2} \tau(x, T) dx \\ + 3 n_E k_B \frac{x_E^2 e^x}{(e^{x_E} - 1)^2} \alpha^2 \frac{\pi}{\omega_E} A e^{-\frac{E_a}{k_B T}} \tag{S6}$$

where $x = \frac{\hbar \omega}{k_B T}$, $k_B$ is Boltzmann constant, $\tau(x, T)$ is spectral relaxation time, $v_j$ is the group velocity of acoustic branch $j$. $\theta_{cj}$ is similar to the Debye temperature of branch $j$, defined as $\theta_{cj} = \hbar \omega_{cj} / k_B$. For cross-plane thermal transport, $\alpha$ is the taken as the distance between the atomic layers. Table. S3 shows the related parameters for calculating the cross-plane thermal conductivity of AV$_3$Sb$_5$ using the hopping model, where $v_j$ is derived from the phonon dispersion calculated by DFT[8] and $\tau(x, T)$ is taken from Cahill-Pohl model[14]. The $A$ defaults to 1 while $n_E$, $\omega_E$ and $E_a$ are fitting parameters, with best-fit values listed Table. S3. The best-fit $n_E$ is found very close to $n$, suggesting that the cross-plane thermal transport is dominated by the hopping mechanism, due to the small density of states of the low-frequency propagating modes. Figure. S4 shows that the hopping model proposed in this work accurately captures the glass-like temperature dependence of $\kappa_z$. The temperature-dependent $\kappa_z$ by Cahill-Pohl model[14] and the Callaway model[15] are also included for comparison.

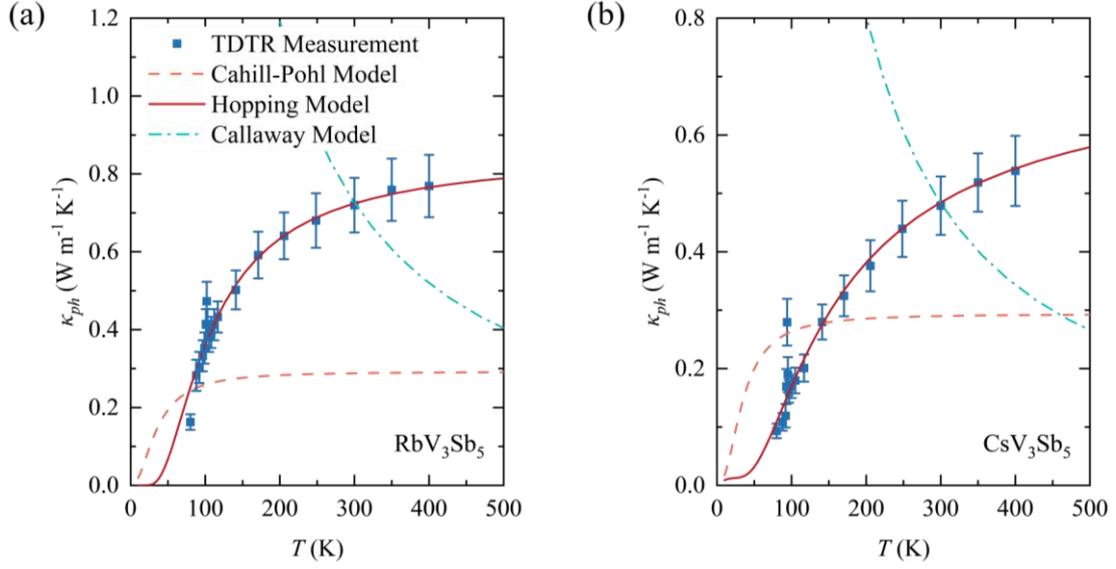

Fig. S4. Data fitting of TDTR-measured $\kappa_{ph}$ along the cross-plane direction of (a) RbV$_3$Sb$_5$ and (b) CsV$_3$Sb$_5$ from 80 K to 400 K, using the hopping model, Cahill-Pohl model, and Callaway model.

Table. S3 Parameters of the hopping model fitting the TDTR-measured $\kappa_{ph}$ of AV$_3$Sb$_5$

| Sample | $v_j$ | $n$ | $n_E$ | $\hbar\omega_E/k_B$ | $\alpha$ | $A$ | $E_a/k_B$ |
|---|---|---|---|---|---|---|---|
| | m s$^{-1}$ | (10$^{28}$ m$^{-3}$) | (10$^{28}$ m$^{-3}$) | (K) | (Å) | | (K) |
| RbV$_3$Sb$_5$ | 2200 ($l$) 1210 ($t$) | 3.83 | 3.80 | 257 | 2.27 | 1 | 35 |
| CsV$_3$Sb$_5$ | 1960 ($l$) 1420 ($t$) | 3.64 | 3.62 | 213 | 2.34 | 1 | 117 |